\documentclass[aps,prl,twocolumn,groupedaddress,showpacs]{revtex4-1}

\usepackage{graphicx} 
\begin{document}


\title{Optical phonon drag and variable range hopping mechanisms of thermoelectric power generation in charge density wave system o-TaS$_{3}$}


\author{D. Stare\v{s}ini\'{c}}
\email[]{damirs@ifs.hr}
\author{M. O\v{c}ko}
\author{K. Biljakovi\'{c}}
\author{D. Dominko}
\affiliation{Institute of physics, Bijeni\v{c}ka cesta 46, P.O.B. 304, HR-10001 Zagreb, Croatia}


\date{\today}

\begin{abstract}
We have measured the thermoelectric power of charge density wave system $o$-TaS$_{3}$ in a wide temperature range between 10 K and 300 K. All features characteristic for charge density wave systems have been observed; strong increase below charge density transition temperature $T_{P}$, peak at $T_{M}\sim$1/2-1/3 $T_{P}$ and change of sign at lower temperatures. These features can be explained by two contributions; free carriers drag by the optical phonons with frequencies close to the density wave gap value in the peak region and above, and hopping of localized carriers of opposite charge to free carriers at low temperatures. We argue that these are common mechanisms of thermoelectric power generation in all charge density wave systems.
\end{abstract}


\maketitle


Charge density wave (CDW) ground state is the best known case of electronic crystals emerging from strong interplay of charge and lattice degrees of freedom in correlated low-dimensional electronic system. For many decades they present a host of unusual phenomena and the ideas that explain these phenomena span much of the contemporary condensed matter physics \cite{MonAP2012,GrunerRMP1988,Gruner1994}. Among exceptional properties of CDW systems are giant dielectric constant, nonlinear transport, pulse-duration memory effects, unusual electro-mechanical and thermoelectric properties, all of conceptual importance in understanding systems with various collective ground states. New technologies require use of materials at smaller and smaller scales, which brings challenges not only to the manufacturing, but also to the understanding of their properties which could change drastically by lowering sample size. Therefore the CDW phenomenology becomes especially attractive as it originates from inherent low dimensionality of CDW systems and concomitant very strong electron-phonon (e-ph) coupling.

The strong increase of the thermoelectric power (TEP) below the transition temperature $T_{P}$ in CDW systems \cite{Shchegolev1972,Chaikin1973,Green1976,Allgeyer1982,Fisher1983,Johnston1983,Stokes1984,Higgs1985,Dian1988,Preob1989,Smontara1989,Almeida1991,Tian1992,Smontara1992,Demishev1998}, with maximum values entering into the range of mV/K, is one peculiarity more which remains unexplained. It was only colloquially referred to as the possible phonon drag contribution \cite{Chaikin1973,Allgeyer1982,Tian1992,Demishev1998}. The softening (Kohn) phonons involved in CDW formation are expected to drag the free carriers \cite{Tian1992}, however, it was clearly demonstrated that, once condensed in CDW state, these phonons do not carry heat \cite{Stokes1984} and cannot contribute to TEP. As phonon drag may produce a considerable contribution to TEP in general, the deeper understanding how it works in systems with strong e-ph interaction which finally causes various collective states is of great importance.

Phonon drag was definitely shown to be important in the TEP of high-temperature superconductors (HTSC). An anomalous, positive component to TEP, observed in YBaCuO for $T<$160 K, was attributed to freeze-out of carrier-phonon Umklapp processes involving holes in the CuO$_{2}$ planes and optical-mode phonons \cite{Cohn1991}. An universal dependence of the TEP on hole concentration of HTSC cuprates was also found \cite{Obertelli1992}. The direct phonon-drag TEP correlation to $T_{c}$ in HTSC was seen as an increase of the phonon-drag contribution with $T_{c}$ consistent with strong-coupling theory \cite{Jones1995}. Finally, TEP of ZrB${12}$ superconductor has been explained by the contribution from the optical phonon associated with the vibrations of the weakly bound Zr atoms in boron cages \cite{Glushkov2006}.

In this Letter we argue that thermoelectric properties of CDW systems show generic temperature dependence featuring the \textit{optical phonon drag} peak at temperature $T_{M}<T_{P}/2$ and change of sign on further cooling indicating a new channel of \textit{hopping conductivity}. Our consistent interpretation of the integral temperature behavior of TEP in CDW systems is demonstrated on thoroughly investigated system $o$-TaS$_{3}$ \cite{Allgeyer1982,Fisher1983,Johnston1983,Stokes1984,Higgs1985,Dian1988,Preob1989} with new measurements in a wide temperature range.

We used the standard differential method to investigate several samples of 6 mm long fibers of 10-20 $\mu$m diameter synthesized by H. Berger at Ecole Polytechnique Fédérale in Lausanne. Measured values of TEP are more than two times larger than the values reported in earlier investigations indicating higher quality of our samples.

\begin{figure}
\includegraphics[width=8.6cm]{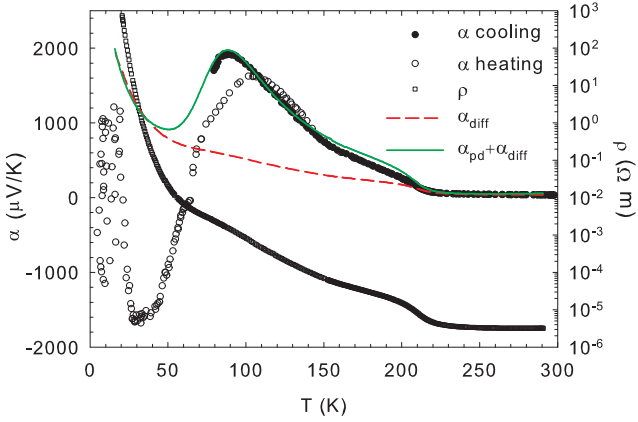}%
\caption{\label{Figure1}
(color online) Thermoelectric power $\alpha$ and resistivity $\rho$ of $o$-TaS$_{3}$ measured along the conducting chains in cooling (full symbols) and heating regime (empty symbols). The dashed red line represents the diffusion contribution to TEP obtained from Eq. (\ref{Eq1}) and the full green line represents the combined diffusion contribution and phonon-drag peak obtained from Eq. (\ref{Eq2}).
}\end{figure}

Fig. \ref{Figure1} shows the temperature dependence of TEP $\alpha$ and resistivity $\rho$ of $o$-TaS$_{3}$. At high temperatures TEP is positive, consistent with the predominantly hole-like carriers \cite{Latyshev1983}. Typical features of TEP in CDW systems are observed, which include strong increase below $T_{P}$=217 K peaked at $T_{M}\sim$ 90 K ($T_{P}$/2-$T_{P}$/3) and change of sign at $T^{*}\sim$60 K with almost equally deep minimum at $T_{m}\sim$30 K. As the sample resistance exceeds G$\Omega$, measurements below 20 K were unstable, so there is uncertainty in the observed decrease of the absolute value of TEP and further change of sign. 

We consider first the high temperature range with positive TEP. Increase of TEP below $T_{P}$ is usually attributed to the semiconducting nature of CDW. The carrier diffusion TEP in semiconductors \cite{Johnson1953} cannot exceed the value 

\begin{equation} \label{Eq1} 
\alpha _{\mathrm{diff}} =\frac{k_{B} }{e} \cdot \frac{\Delta }{k_{B} T} +\alpha _{0}  
\end{equation} 

where $\Delta$ is the effective activation energy for the conductivity and $\alpha_{0}$ is a constant of the order of $k_{B}/e\approx$86 mV/K, with $k_{B}$ the Boltzmann constant and $e$ elementary charge. With $\Delta(T)$ appropriately estimated from $\rho(T)$ data \cite{Nguyen1994} and $\alpha_{0}\approx$-1.5$\cdot k_{B}/e$ set to match the $T>T_{P}$ value, this contribution, represented by dashed red line in Fig. \ref{Figure1}, cannot account for the measured increase of TEP below $T_{P}$.

We have therefore to consider other heat transport mechanisms, such as phonons, which can drag the carriers and produce increased TEP. We use a simplified expression for the phonon drag TEP $\alpha_{pd}$ which assumes energy independent scattering rates as following \cite{Guenault1971,Chaikin1991}:

\begin{equation} \label{Eq2} 
\alpha_{\mathrm{pd}} =\frac{C_{\mathrm{ph}}}{n_{fc}\cdot e} \left(\frac{\Gamma _{\mathrm{e-ph}}}{\Gamma _{\mathrm{e-ph}}+\Gamma _{\mathrm{other}}} \right),  
\end{equation} 

where $C_{\mathrm{ph}}$ is the phonon specific heat, $n_{\mathrm{fc}}$ is the density of free charge carriers and $\Gamma_{\mathrm{e-ph}}$ and $\Gamma_{\mathrm{other}}$ are the scattering rates of phonon subsystem by charge carriers and by other scattering channels respectively.

Both acoustic and optical phonons can contribute to TEP \cite{Cohn1991,Glushkov2006}, as real optical phonons, unlike the ideal Einstein modes, are not dispersionless and they can contribute significantly to the heat transport \cite{Hess2004}. However, the contribution of a particular phonon mode to TEP will be negligible unless scattering by the free carriers is important, i.e. $\Gamma_{\mathrm{e-ph}}/(\Gamma_{\mathrm{e-ph}}+\Gamma_{\mathrm{other}})\equiv \Gamma_{\mathrm{rel}}\approx$1.

Raman measurements \cite{Tsang1978,Sugai1984,Sood1985} in $o$-TaS$_{3}$ detected three groups of optical phonons near 280 cm$^{-1}$, 380 cm$^{-1}$ and 500 cm$^{-1}$ that are strongly coupled to the free carriers, as their width ($\Gamma$) increases several times at 175 K, 160 K and 80 K respectively, temperatures where their frequency matches the value of the temperature dependent half-gap $\Delta(T)$ \cite{Tsang1978}. At these temperatures the free carrier scattering becomes dominant and  $\Gamma_{\mathrm{e-ph}}>\Gamma_{\mathrm{other}}$. The mechanism of coupling with electrons was not elucidated, however the bolometric measurements indicated strong coupling of free carriers with these modes \cite{Herr1986}, presumably due to the large transverse dipole moment of conduction band states.

At sufficiently high temperatures, where $C_{\mathrm{ph}}$ and $\Gamma_{\mathrm{rel}}$ are nearly constant, $\alpha_{\mathrm{pd}}$ will be inversely proportional to the free carrier density, and therefore roughly proportional to $\rho$. $\alpha_{\mathrm{pd}}(T)$ is then expected to follow the complex temperature dependence of $\rho(T)$ around $T_{P}$ and increase at lower temperatures as evidenced in Fig. \ref{Figure1}.

On the other hand, at sufficiently low temperatures $C_{\mathrm{ph}}$ of the optical phonon will decrease exponentially and, if its frequency is higher than $\Delta$, $\alpha_{\mathrm{pd}}$ will decrease as the temperature is decreased, leading to the maximum in TEP. Experimentally observed $\Gamma_{\mathrm{e-ph}}$ decreases as well, reducing $\Gamma_{\mathrm{rel}}$ and the phonon drag contribution to TEP even further.

In order to model the phonon drag contribution according to Eq. (\ref{Eq2}) we assume that $\Gamma_{\mathrm{rel}}$ is constant and close to 1, which would be acceptable approximation for the experimentally observed Raman mode near 500 cm$^{-1}$ above 80 K. Based on the simplified expression for the free carrier conductivity $\sigma=n_{\mathrm{fc}} e \mu$, where $\mu$ represents mobility, $n_{\mathrm{fc}}(T)$ is estimated as $n_{\mathrm{fc}}(T)=n_{0}\cdot \sigma(T)/\sigma(300 K)$ from the measured temperature dependence of the conductivity $\sigma(T)=1/\rho(T)$, assuming that $\mu$ varies slowly with temperature. $n_{0}\approx 10^{22} e/cm^{3}$ is the free carrier density in the high temperature metallic state \cite{Ong1982}.

$C_{\mathrm{ph}}(T)$ is estimated from the Einstein formula for the heat capacity which is a fair approximation for optical phonons of frequency $\omega_{E}$. As the unit cell of $o$-TaS$_{3}$ consists of 24 chains (or formula units) and the main optical modes come from the prismatic structure of single TaS$_{3}$ chain, the basic degeneracy of the single mode is 24 \cite{Sugai1984}.

We have varied only the optical phonon frequency $\omega_{E}$ and obtained quite satisfactory fit for the peak in TEP with $\alpha_{\mathrm{pd}}+\alpha_{\mathrm{diff}}$, as shown in Fig. \ref{Figure1} (solid green line), for $\omega_{E}\approx$480 cm$^{-1}$. This frequency is close to the 500 cm$^{-1}$ optical phonons for which a strong increase of the line width has been observed at 80 K \cite{Sugai1984}, i.e. below the position of the peak in TEP. The corresponding energy $\hbar \omega_{E}/k_{B}$=690 K is also  close to the effective activation energy $\Delta\approx$750 K for the conductivity. This justifies some of the most important assumptions we have made in the calculation.

\begin{figure}
\includegraphics[width=8.6cm]{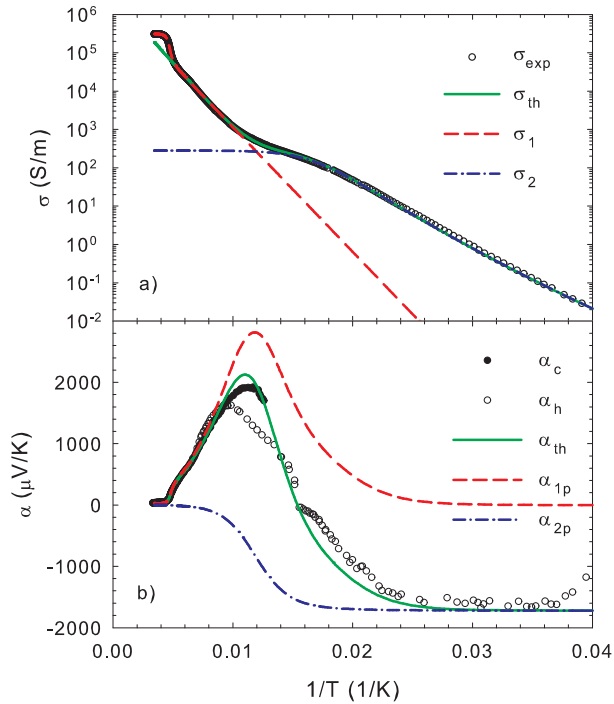}%
\caption{\label{Figure2}
(color online) a) electrical conductivity of $o$-TaS$_{3}$ measured (symbols) and decomposed in two contributions from free (red dashed line) and hopping (blue dash-dotted line) carriers as described in the text. The green solid line is the fit to Eq. (\ref{Eq4}. b) TEP of $o$-TaS${}_{3}$ measured (symbols) and estimated from Eq. (\ref{Eq3}) (green solid line) with partial contributions (Eq. (\ref{Eq5})) from the free carriers (red dashed line), including carrier diffusion and phonon drag, and hopping carriers (blue dash-dotted line).
}\end{figure}

We consider now the low temperature regime below the maximum of TEP, which is dominated by the change of sign at $T^{*}$. Apparently, it should be attributed to the new transport channel with the opposite carrier charge. This low temperature behavior can be understood in the frame of new generic phase diagram of DW systems in general (including also SDW ground state) with new "density wave glass" ground state established below a finite temperature $T_{g}\sim T_{P}/4-T_{P}/5$ \cite{Staresinic2002,Biljakovic2009}. Below this temperature the free carrier contribution to the conductivity continues to decrease with the same activation energy \cite{Latyshev1983,ZZ2006}, but another channel with lower apparent activation energy $E_{a}=200-400$ K ($\sim T_{P}$) becomes dominant in electrical transport \cite{Zhilinskii1983}, as demonstrated in Fig. \ref{Figure2}a. 

Several microscopic models of this low temperature CDW state predict localized carriers of opposite charge \cite{Artemenko1995,Artemenko2005} (i.e. the electrons in the case of $o$-TaS$_{3}$) which could explain the observed change of sign in TEP. Moreover, the hopping mechanism of the corresponding conductivity can lead to the temperature independent TEP, as shown in \cite{Burns1985} for the case of variable range hopping (VRH) in 1D Mott and Coulomb gap (Efros-Shklovskii - ES) models, which is a reasonable approximation for the nearly constant negative TEP of -20 $k_{B}/e$ observed below 50 K.

With two transport channels the overall TEP can be obtained from
\begin{equation} \label{Eq3} 
\alpha _{th} =\frac{\alpha _{1} \sigma _{1} +\alpha _{2} \sigma _{2} }{\sigma _{1} +\sigma _{2} }  
\end{equation} 
where $\alpha_{1}$ and $\sigma_{1}$ represent the free carrier contribution to TEP and conductivity, and $\alpha_{2}$ and $\sigma_{2}$ the hopping contribution.
 
\noindent Taking into account that $\sigma_{1}$ shoud decrease in the activated manner in the entire temperature range below $T_{P}$ \cite{Latyshev1983,ZZ2006}, $\sigma_{2}$ can be obtained numerically. However, in order to simplify further calculations we decompose the temperature dependence of conductivity $\sigma(T)$ below 150 K as follows:

\begin{equation} \label{Eq4} 
\sigma _{th} \left(T\right)=\sigma _{1} +\sigma _{2} =\sigma _{01} e^{-\frac{\Delta }{k_{B} T} } +\frac{\sigma _{02} }{1+be^{\left(\frac{T_{0} }{T} \right)^{\frac{1}{2} } } }  
\end{equation} 

represented by green line in Fig. \ref{Figure2}a. Above 150 K $\sigma_{1}$ is taken directly from the experiment. Resulting free carrier and hopping carrier conductivities are represented by dashed red ($\sigma_{1}$) and dash-dotted blue ($\sigma_{2}$) lines respectively in Fig. \ref{Figure2}a.

The value of $\Delta$=750 K is typical for $o$-TaS$_{3}$. The analytical expression for $\sigma_{2}$ conforms to the 1D Mott or the ES VRH model below 50 K. At higher temperatures it mimics the numerical VRH results of \cite{Pasveer2005} where a constant conductivity is obtained above sufficiently high temperature. The parameter $b=2.2\cdot10^{-8}$ determines the crossover temperature of 60 K. VRH model has not been previously applied to the in-chain conductivity of $o$-TaS$_{3}$ above 20 K, however, the fit is reasonable and the value of $T_{0}=18.4\cdot10^{3}$ K, is very close to $T_{0}$ obtained for VRH transversal conductivity of $o$-TaS$_{3}$ \cite{Pokrovskii2005}.

The free carrier contribution to TEP $\alpha_{1}$ includes the diffusion and phonon drag, however, with $\Delta(T)$ in Eq. (\ref{Eq1}) and $n_{\mathrm{fc}}(T)$ in Eq. (\ref{Eq2}) estimated from $\sigma_{1}$. VRH contribution to TEP, $\alpha_{2}$=20$\cdot k_{B}/e$, is taken directly from experimental results.

Two adjustable parameters that we have varied in order to obtain a good fit to experimental data are the optical phonon frequency $\omega_{E}$ and the scattering term $\Gamma_{\mathrm{rel}}$ in Eq. (\ref{Eq2}). TEP evaluated for $\Gamma_{\mathrm{rel}}\approx$0.6 and $\omega_{E}\approx$380 cm$^{-1}$ is presented in Fig. \ref{Figure2}b (green solid line), together with partial contributions of free carriers ($\alpha_{1\mathrm{p}}$) and hopping ($\alpha_{2\mathrm{p}}$):

\begin{equation} \label{Eq5} 
\begin{array}{lr} {\alpha _{1\mathrm{p}} =\frac{\alpha _{1} \sigma _{1} }{\sigma _{1} +\sigma _{2} } } & {\alpha _{2\mathrm{p}} =\frac{\alpha _{2} \sigma _{2} }{\sigma _{1} +\sigma _{2} } } \end{array} 
\end{equation} 

The frequency $\omega_{E}$ is within the range of the optical phonons observed by Raman spectroscopy \cite{Sugai1984} to be strongly scattered by the free carriers. For particular 380 cm$^{-1}$ phonon the free carrier scattering actually becomes dominant at temperatures higher than $T_{M}$, which is reflected in the reduced value of $\Gamma_{\mathrm{rel}}$. However, these quantitative considerations should not be taken too strictly, as within our simple model optical phonon parameters represent the average contribution of all affected phonons.

Anomalous increase of TEP below $T_{P}$ and subsequent maximum are common features of CDW systems. In  proposed optical phonon drag mechanism of TEP, these features are related to the phonons with frequencies crossing the CDW gap energy. In Table 1. we compare the $\Delta$, $T_{P}$ and $T_{M}$ values of several CDW systems for which TEP has been measured in sufficiently broad temperature range. In respect to the wide variation of $T_{P}$ and $\Delta$, as well as 2-fold variation of the $\Delta/T_{P}$ ratio, the measure of the coupling strength, $T_{M}$ scales reasonably well with the value of $\Delta$. It is a good indication that the same universal mechanism with $\Delta$ as the relevant energy scale, such as proposed optical phonon drag, is responsible for TEP in CDW systems below $T_{P}$.

\begin{table}
\caption{\label{Table}
Comparative values for the CDW gap $\Delta$, CDW transition temperature $T_{P}$ and the temperature of the maximum in TEP $T_{M }$ and their ratios for various CDW systems, $o$-TaS$_{3}$ (this work), K$_{0.3}$MoO$_{3}$ \cite{Almeida1991}, TTF-TCNQ \cite{Chaikin1973}, HMTTF-TCNQ \cite{Green1976}, BDT-TTF \cite{Demishev1998}, KCP \cite{Shchegolev1972}, (TaSe$_{4}$)$_{2}$I \cite{Smontara1989}, (NbSe$_{4}$)$_{10/3}$I\cite{Smontara1992}.
}
\begin{ruledtabular}
\begin{tabular}{lllllll}
 & $\Delta$ (K) & $T_{P}$ (K) & $T_{M}$ (K) & $\Delta/T_{P}$ & $T_{P}/T_{M}$ & $\Delta/T_{M}$ \\ 
$o$-TaS$_{3}$ & 750 & 210 & 90  & 3.6 & 2.3 & 8.3 \\ 
K$_{0.3}$MoO$_{3}$& 600 & 175 & 60 & 3.4 & 2.9 & 10 \\  
TTF-TCNQ& 220 & 53 & 20 & 4.2 & 2.7 & 11 \\  
HMTTF-TCNQ& 250 & 50 & 25 & 5 & 2.0 & 10 \\  
BDT-TTF& 500 & 150 & 40 & 3.3 & 3.8 & 12.5 \\ 
KCP& 500  & 130 & 50 & 3.8 & 2.6 & 10 \\  
(TaSe$_{4}$)$_{2}$I& 1600 & 260 & 130 & 6.2 & 2.0 & 12.3 \\  
(NbSe$_{4}$)$_{10/3}$I& 1850 & 280 & 190 & 6.6 & 1.5 & 9.7 \\  
\end{tabular}\end{ruledtabular}
\end{table}

The change of sign in TEP below $T_{M}$ is related in our model to the additional channel of conductivity, attributed to the hopping mechanism. Therefore the temperature $T^{*}$ at which TEP changes sign is not related to $\Delta$. Moreover, if the measurements in the literature are performed at sufficiently low temperatures to determine $T^{*}$, they have not been low enough to permit the determination of low temperature value of TEP required in the fit.

Our interpretation of TEP of $o$-TaS$_{3}$ in a wide temperature range relies of two important presumptions, the free carrier drag by optical phonons and hopping contribution to the conductivity at low temperatures. Although these have not been considered so far in the theory of CDW systems, except for hopping in \cite{Artemenko2005}, they are formulated according to the experimental results and the corresponding fits are obtained with the parameters extracted from these results.

There is, however, remaining question about the origin of the strong coupling of optical phonons with free carriers seen in the Raman scattering. It occurs at the phonon energies close to $\Delta$, which is only half of the optical gap, and therefore does not correspond to the simple creation of electron-hole pairs. So far, there is no mechanism that can explain it, which is appealing for further theoretical consideration.

In summary, we have successfully described the thermoelectric power of CDW system $o$-TaS$_{3}$ in a wide temperature range with two contributions corresponding to the optical phonon drag of free carriers excited over the gap and the hopping of localized carriers of charge opposite to free carriers. Comparison with other CDW systems suggests that these are common mechanisms of TEP in all of them.

The authors thank Pierre Monceau and Ivan Kupèiæ for useful discussions and suggestions. This study was supported by Croatian MSES projects 035-0352827-2841 and 035- 0352827-2842.

%



\bibliography{CDW_TEP}

\end{document}